\documentclass[11pt]{article}
\usepackage[totalwidth=385pt,totalheight=584pt]{geometry}
\usepackage{amsmath,amssymb,epsf,graphics,graphicx}
\usepackage{psfrag}
\newcommand{\CM}{{\cal M}}

\newcommand{\ZZ}{\mathbb{Z}}

\newcommand\blank[1]{}

\newcommand{\fract}[2]{{\textstyle\frac{#1}{#2}}}

\renewcommand{\hat}{\widehat}
\newcommand\eq{\begin{equation}}
\newcommand\en{\end{equation}}
\newcommand\bea{\begin{eqnarray}}
\newcommand\eea{\end{eqnarray}}

\newcommand\ba{\(\begin{array}}
\newcommand\ea{\end{array}\)}
\newcommand{\resection}[1]{\setcounter{equation}{0}\section{#1}}

\newcommand{\NN}{{\mathbb N}}

\newcommand\Rth{{\mathbb R}}
\newcommand\p{{\sf a}}
\newcommand\q{{\sf b}}
\newcommand\pp{{\sf P}}

\begin{document}
\begin{titlepage}
\vskip 0.5cm
\begin{flushright}
NI07080 \\
DCPT-07/69 \\
UKC/IMS/07/019 \\
arXiv:0712.2010 \\
December 2007\\
\end{flushright}
\vskip 1.2cm
\begin{center}
{\Large{\bf On the ODE/IM correspondence for minimal models}}
\end{center}
\vskip 0.8cm \centerline{Patrick Dorey$^{1,2}$,
Clare Dunning$^3$, Ferdinando  Gliozzi$^4$
and Roberto Tateo$^4$} \vskip 0.9cm
\centerline{${}^1$\sl\small Dept.\ of Mathematical Sciences,
Durham University,} \centerline{\sl\small Durham DH1 3LE, UK}
\vskip 0.3cm \centerline{${}^{2}$\sl\small Isaac Newton Institute,
20 Clarkson Road, Cambridge CB3 0EH, UK}
\vskip 0.3cm \centerline{${}^{3}$\sl\small IMSAS, University of
Kent, Canterbury CT2 7NF, UK}
\vskip 0.3cm \centerline{${}^{4}$\sl\small Dip.\ di Fisica Teorica
and INFN, Universit\`a di Torino,} \centerline{\sl\small Via P.\
Giuria 1, 10125 Torino, Italy}
\vskip 0.2cm \centerline{E-mails:}
\centerline{p.e.dorey@durham.ac.uk, t.c.dunning@kent.ac.uk,}
\centerline{ gliozzi@to.infn.it,
tateo@to.infn.it}

\vskip 1.25cm
\begin{abstract}
\noindent
Within the framework of the ODE/IM correspondence, we show that the
minimal conformal field theories with $c<1$ emerge naturally from
the monodromy properties of certain families of ordinary
differential equations.

\end{abstract}
\end{titlepage}
\setcounter{footnote}{0}
\def\thefootnote{\fnsymbol{footnote}}
%
\resection{Introduction}
This note is about a unifying  programme, the ODE/IM correspondence,
which links two-dimensional conformal field theories and
integrable
models  to the spectral theory of ordinary differential equations.
The first instance of this correspondence~\cite{Dorey:1998pt}
was based on an identity  between the transfer
matrix eigenvalues of certain integrable models
in their conformal limits
\cite{Bazhanov:1994ft,Bazhanov:1996dr}, and the spectral
determinants \cite{Sha,Voros} of second-order ordinary
differential equations.
Since the initial   results of \cite{Dorey:1998pt} and then
\cite{Bazhanov:1998wj},  the ODE/IM correspondence  has been
used in various branches of physics ranging from condensed
matter~\cite{GADP} to PT-symmetric quantum
mechanics~\cite{DDTb}, and from boundary conformal field
theory~\cite{Lukyanov:2003nj} to the study of non-compact sigma
models~\cite{Teschner:2007ng}. It has also been linked
with the geometric Langlands
correspondence~\cite{Feigin:2007mr}. A recent review containing many
more references is \cite{Dorey:2007zx}.

Early examples of the correspondence concerned the ground states of
the integrable lattice models, albeit with possibly twisted boundary
conditions. In the conformal field theory setting this gave access
to primary fields, but not to their descendants. However, in
\cite{Bazhanov:2003ni} Bazhanov, Lukyanov and Zamolodchikov
conjectured that the descendant fields
could be found through relatively-simple generalisations of the
initial differential equation. The new equations were obtained
by modifying the initial potential, which in general has a regular
singularity at zero and  an irregular singularity at infinity,  by
introducing a
level-dependent number of additional regular
singularities in the complex plane, subject to a zero-monodromy
condition around  these extra singularities, though not about the
origin. In this short
note we will show that the minimal $c < 1$ conformal field theories
can equivalently be associated to ODEs governed by a trivial
monodromy property in the {\em whole}\/
 complex plane, including the origin.
A natural quantisation condition on the
coefficient of the regular (Fuchsian) singularity at the origin
emerges, and since this coefficient is related to
the Virasoro vacuum parameter $p$ of
\cite{Bazhanov:1994ft,Bazhanov:1996dr}
this restricts the resulting conformal weights, so as
to match precisely the Kac tables of the minimal
models $\CM_{\p \, \q}$.

\section{The $c<1$ minimal models }
\label{sec1}
We begin with the basic observation of
\cite{Dorey:1998pt,Bazhanov:1998wj},
that the Schr\"odinger equation
\eq
\left(-\frac{d^2}{dx^2}+ (x^{2M} -E)+ \frac{l(l+1)}{x^2} \right)
\psi(x,E,l)=0
\label{first}
\en
is related to conformal field theory. The Stokes
relations associated to (\ref{first}) imply
constraints on its eigenvalues $E \in \{ E_i \}$, given suitable
boundary conditions, which
coincide with the Bethe Ansatz equations (BAEs) for the
twisted six-vertex model in its conformal ($c=1$) limit~(see, for
example, \cite{Dorey:2007zx}). The same BAEs emerge from the
study of $c \le 1$ CFTs in the framework developed  by Bazhanov,
Lukyanov and Zamolodchikov in~\cite{Bazhanov:1994ft,
Bazhanov:1996dr}. In the notation used in~\cite{Bazhanov:2003ni},
equation (\ref{first}) encodes the primary field of a
Virasoro module with  central charge $c$, vacuum parameter $p$ and
highest weight $\Delta$, where
\eq
c(M)= 1- \frac{6 M^2}{M+1}~,~~p=\frac{2l+1}{4M+4}~,~~
\Delta(M,l)=\frac{(2l
+ 1)^2 -4 M^2}{16 (M + 1)}~.
\label{cdelta}
\en
In \cite{Dorey:1999uk} it was observed that for $2M$ rational
and suitable values of $l$, the solutions to (\ref{first}) will all
lie on a finite cover of the complex plane, and that this
translates into a truncation of the fusion hierarchy \cite{KP}
of the associated integrable model. This is of particular interest
because, for
such truncations, the central charges and field contents map to those of
the minimal models with $c<1$ (see, for
example, the discussion in \S3  of \cite{Bazhanov:1996aq}).
However only the simplest case of $2M$ integer and $l(l{+}1)=0$ was
discussed explicitly in \cite{Dorey:1999uk}, in part
because the general monodromies of solutions are hard to unravel
in the presentation (\ref{first}). One of our aims in this note is to
show that a much simpler treatment is possible, from which the Kac
tables of minimal model primary fields emerge in a very natural
fashion.

Working backwards, we first note that, for any two coprime integers
 $\p<\q$,  the ground state of the
minimal model $\CM_{\p \,\q}$
is found by setting
\eq
M+1=\frac{\q}{\p}~,~~
 l+\frac{1}{2}=\frac{1}{\p}~
\en
in (\ref{first}). This corresponds to the
central charge $c_{\p\q}=1-\frac{6}{\p\q}(\q{-}\p)^2$ and
(lowest-possible) conformal weight
$\Delta=\frac{4}{\p\q}(1-(\q{-}\p)^2)$.

We now observe, as in \S 6 of
\cite{Dorey:2004fk}, that
the $l(l{+}1)/x^2$ term in (\ref{first}) can be eliminated, for
this value of $l$,
by the following transformation:
\eq
x= z^{\p/2}~,~~\psi(x, E)=z^{\p/4-1/2} y(z,E)\,.
\label{change}
\en
With a further rescaling $z\to (2/\p)^{2/\q}z$, (\ref{first})
becomes
\eq
\left(-\frac{d^2}{dz^2}+
z^{\p-2} (z^{\q-\p} -\tilde E) \right) y(z,\tilde E)=0
\label{final}
\en
where
\eq
\tilde E = \left(\frac{\p}{2}\right)^{2-2\p/\q}E\,.
\en
Notice that the change of variable has replaced the singular
generalised potential
$P(x)=x^{2\q/\p-2}-E+(1/\p^2-1/4)x^{-2}$, defined on a
multi-sheeted Riemann surface, by a simple polynomial
$W(z)=z^{\p-2} (z^{\q-\p}- \tilde E)$. In particular, any solution
to (\ref{final}) is automatically
single-valued  around $z=0$, and the truncation of the fusion
hierarchy as explained in \S 4 of
\cite{Dorey:1999uk} is made much more transparent.

To see which other primary states in the original model might have
similarly-trivial monodromy, we keep $l$
real with $l+1/2>0$, but otherwise arbitrary, and again perform the
change of variable (\ref{change}). The result is now
\eq
\left(-\frac{d^2}{dz^2}+  \frac{\tilde l  (\tilde l +1)}{z^2}
+ z^{\p-2} (z^{\q-\p} -\tilde E)
 \right) y(z,\tilde E,\tilde l)=0
\label{finale}
\en
where
\eq
2(\tilde{l}+\fract{1}{2})=\p (l+ \fract{1}{2})~.
\label{tildel}
\en
The Fuchsian singularity in (\ref{finale}) at $z=0$ means that the
equation admits a pair of solutions which generally have the power
series expansions
\eq
\chi_1(z)=z^{\lambda_1}\sum_{n=0}^\infty c_n z^n~;~~~~
\chi_2(z)=z^{\lambda_2}\sum_{n=0}^\infty d_n z^n~~,
\label{chichi}
\en
where $\lambda_1=\tilde{l} +1$ and $\lambda_2=-\tilde{l}$
are the two roots of the indicial equation
\eq
\lambda(\lambda-1)-\tilde l(\tilde l+1)=0~.
\en
A general solution to (\ref{finale}) can be expressed as
$y(z,\tilde E,\tilde l)=\sigma\chi_1(z)+\tau \chi_2(z)$, and
we shall demand that the transformed ODE
should be such that, for arbitrary $\tilde E$, the monodromy
of $y(z,\tilde E,\tilde l)$
around $z=0$
is projectively trivial, which is to say that
\eq
y(e^{2 \pi i}z) \propto y(z)\,.
\en
This condition ensures that the
eigenvalues obtained by imposing the simultaneous decay of solutions
in a pair of asymptotic directions at infinity are independent of
the path of analytic continuation between these two directions. We
shall show that the condition imposes the following constraints
on~$\tilde l$:
\begin{itemize}
\item[i)]$2\tilde l+1$ is a positive integer;
\item[ii)] The allowed values of $2\tilde l+1$ are
those integers which cannot be written as $\p\,s+\q\,t$ with $s$ and
$t$ non-negative integers. In other words they form precisely the
set of
 holes of the infinite sequence
\eq
\p\,s+\q\,t~~,~~s\,,t=0,1,2,3\dots~.
\label{sequence}
\en
We  shall call the integers (\ref{sequence}) `representable' and
denote the set of them by $\Rth_{\p\q}$. 
\end{itemize}
As a consequence we shall see that, as $\tilde l$ runs over its
`allowed' values, the  rational numbers
\eq
\Delta_{\tilde l}=\Delta(M,l)|_{M=\q/\p{-}1,l=l(\tilde l)}
=\frac{(2\tilde l+1)^2-(\p-\q)^2}{4\p\q}
\label{deltal}
\en
precisely reproduce the set of conformal weights of the primary
states lying in the Kac table of the minimal model $\CM_{\p \,\q}$.
Figures \ref{holes1} and \ref{holes2} illustrate the story for the
Ising and Yang-Lee cases.

\[
\begin{array}{c}
\!\!\!
\includegraphics[width=0.5\linewidth]{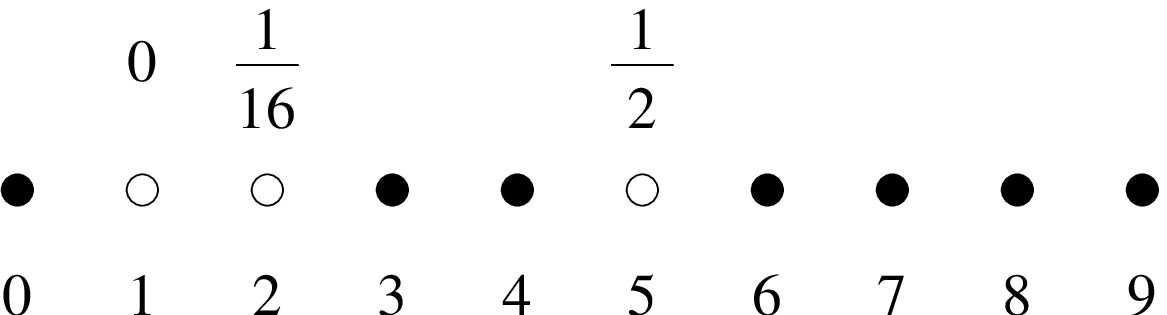}
\\[11pt]
\parbox{0.95\linewidth}{
\small
{\small Figure \ref{holes1}:
 The holes (open circles) in the infinite sequence of
integers defined in (\ref{sequence}) for
the critical Ising model $\CM_{3\,4}$. The holes are
at $1$, $2$ and $5$; the resulting conformal weights according to
(\ref{deltal}) are also shown, matching the primary field content of
the Ising model. }}
\end{array}
\refstepcounter{figure}
\label{holes1}
\]

\[
\begin{array}{c}
\!\!\!
\includegraphics[width=0.5\linewidth]{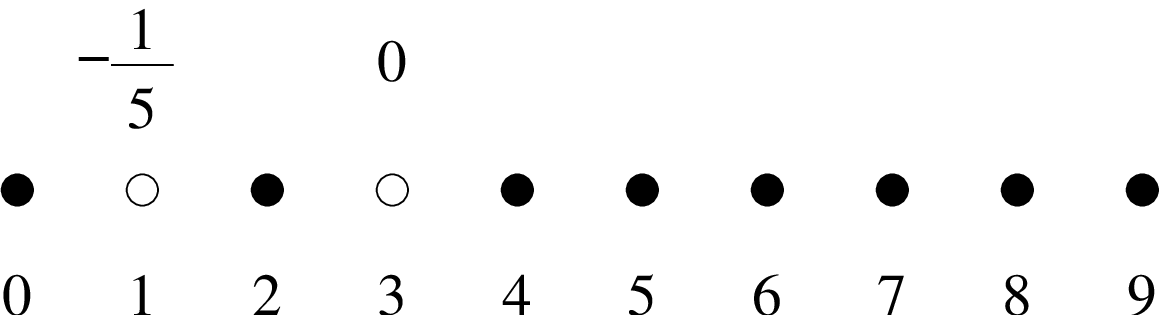}
\\[11pt]
\parbox{0.95\linewidth}{
\small
{\small Figure \ref{holes2}: Holes for the Lee-Yang model
$\CM_{2\,5}$\,, at $1$ and $3$. Notation as in Figure~\ref{holes1}.
 }}
\end{array}
\refstepcounter{figure}
\label{holes2}
\]

To establish these claims we first note that the
requirement that the general solution $y(z)$ be projectively trivial
means that $\chi_1(z)$ and $\chi_2(z)$ must have the same monodromy,
which implies that
the two roots of the indicial equation must differ by an integer:
\eq
\lambda_1-\lambda_2=2\tilde l+1\in\NN\,.
\en
This in turn restricts $\tilde l$ to be an integer or half
integer, so that, na\"ively, the allowed solutions are even or odd
under a $2\pi i$ rotation around $z=0$.
However it is well known that in such a circumstance, while $\chi_1(z)$
keeps its power series expansion (\ref{chichi}), $\chi_2(z)$
generally acquires a logarithmic contribution:
\eq
\chi_2(z)= D\chi_1(z)\log(z)+\frac1{z^{\tilde l}}
\sum_{n=0}^\infty d_nz^n~.
\label{chilog}
\en
Unless $D=0$, this will spoil the projectively trivial monodromy of
$y(z)$. We now show that $D=0$ if and only if $2\tilde
l+1$ obeys the constraint ii). In fact the logarithmic term is only
absent when the recursion relations for the $d_n$'s with $D=0$ admit
a solution.
These relations are
\eq
n\,(n-2\tilde l-1)\,d_n=d_{n-\q}-\tilde E\,d_{n-\p}
\label{recursion}
\en
with the initial conditions $d_0=1$, $d_{m<0}=0$.

Consider first the situation when $2 \tilde l+1 \notin \Rth_{\p\q}$.
Then, starting from the given initial conditions, the recursion
relation (\ref{recursion}) generates a solution of the form
\eq
\chi_2(z)= \frac{1}{z^{\tilde l}} \sum_{n=0}^{\infty} d_{n}  z^{n}
\label{ser2}
\en
where the only nonzero $d_n$'s are those 
for which the label $n$ lies in the set $\Rth_{\p\q}$.
Given that $2\tilde l+1\notin \Rth_{\p\q}$, for these values of $n$
the factor $n(n-2\tilde l-1)$
on the LHS of (\ref{recursion}) is never
zero, and hence this procedure is well-defined.

If instead $2\tilde l+1\in\Rth_{\p\q}$, then equation (\ref{recursion})
taken at $n=2\tilde l+1$
yields the additional condition
\eq
  \tilde E\,d_{2 \tilde l+1 -\p}-d_{2 \tilde l +1 -\q}=0~,
\label{addcond}
\en
which is inconsistent for generic $\tilde E$, and so the
logarithmic term is required\footnote[3]{ \label{f1} The reader might
wonder whether
the LHS of (\ref{addcond}) could vanish identically as a result of
an exceptional cancellation between the two terms. This can be ruled
out by observing that the monodromy property in $z$ is unaltered by
the change of variable $z
\rightarrow  \beta^{1/\q}z$. This rescaling leads to a  more general
version of
(\ref{recursion}): $n\,(n-2\tilde l-1)\,d_n=\alpha
d_{n-\p}+\beta d_{n-\q}$ with $\alpha=-\tilde E
\beta^{\p/\q}$.
 Then for the particular choice
$\alpha=(-1)^{\p+1}$, $\beta=(-1)^{\q+1}$, one can see that
 $\alpha d_{n-\p}$ and $\beta d_{n-\q}$ must have the same sign for
 $n \le 2\tilde l +1$; 
hence, they cannot cancel identically.}.

Given the characterisation
(\ref{sequence}) of $\Rth_{\p\q}$,
the set $\ZZ^+$ of
non-negative integers can be written as a disjoint union
\eq
\ZZ^+=\Rth_{\p\q}\cup\NN_{\p\q}
\en
where $\NN_{\p\q}$ is the set of `nonrepresentable' integers.
If the coprime integers $\p$ and $\q$ are larger than 1 then
$\NN_{\p\q}$ is non-empty; in fact
$|\NN_{\p\q}|=\frac{1}{2}(\p{-}1)(\q{-}1)$, a result which goes back
to Sylvester \cite{sylv}.

To characterise $\NN_{\p\q}$ more precisely,
we start with the fact that given two
coprime integers $\p$ and $\q$,  any integer $n$ can be written as
\eq
n=\p s_0+\q t_0\,,\quad s_0,t_0\in\ZZ~.
\label{euclid}
\en
This is a classical result of number theory. An intuitive proof
uses Euclid's algorithm  for the  greatest common divisor
of two integers. Alternatively one can invoke the Euler totient
function $\varphi$\footnote[4]{$\varphi(\q)$ denotes  the number of
coprimes with $\q$ in the set $1,2,\dots,\q$. } and the following
theorem~(see, for example,~\cite{DedDir})
\eq
\p^{\varphi(\q)} \equiv1~~(\mbox{mod}~\q)~,
\label{Euler1}
\en
which implies that $\p^{\varphi(\q)} =1+h\q$ for some integer $h$.
This immediately yields the solution $s_0= n\p^{\varphi(\q)-1}$,
$t_0=-n h$. However, for any given $n$, this is not the only
possibility. More precisely, we have $\p s_0+\q t_0=\p s+\q t$ for
some other pair of integers $(s,t)$ if
and only if $s=s_0+\q k$, $t=t_0-\p k$ for some $k\in\ZZ$. This is
easily proved: rearrange (\ref{euclid}) as
\eq
\p(s_0-s)=\q(t-t_0)~.
\label{reuclid}
\en
Since $(\p,\q)=1$, $\q$ must be a factor of $s_0-s$ and so $s_0-s=-\q
k$ for some $k$. Dividing (\ref{reuclid}) by $\q$ then shows that
$t_0-t=\p k$, as required. Hence the possible representatives for
each integer $n$ constitute the line of points $(s,t)=(s_0+\q
k,t_0-\p k)$, $k\in\ZZ$\,. For $n$ to be a {\em positive}\/ integer,
$\p s+\q t>0$, or $s>-\q t/\p$.  If none of these points has both
coordinates non-negative, then the corresponding $n$ will be in
$\NN_{\p \q}$. To keep $t$ non-negative while making $s$ as large as
possible, we shift $t$ by a multiple of $\p$ so that $0\le t<\p$. If
$s$ is still negative, then $n$ will be in $\NN_{\p \q}$. The
numbers we want are therefore represented by the points
\eq
\{(s,t)\,,~0\le t<\p, -\q t/\p <s\le -1\}~.
\en
Negating $t$, the allowed (trivial monodromy)
values of $2\tilde l+1$ are therefore
\eq
2\tilde l+1= \p s-\q t\,,~~~0\le t<\p,~ 1\le s < \q t/\p.
\en
Figure \ref{frob} illustrates the argument.

\[
\begin{array}{c}
\!\!\!\includegraphics[width=0.6\linewidth]{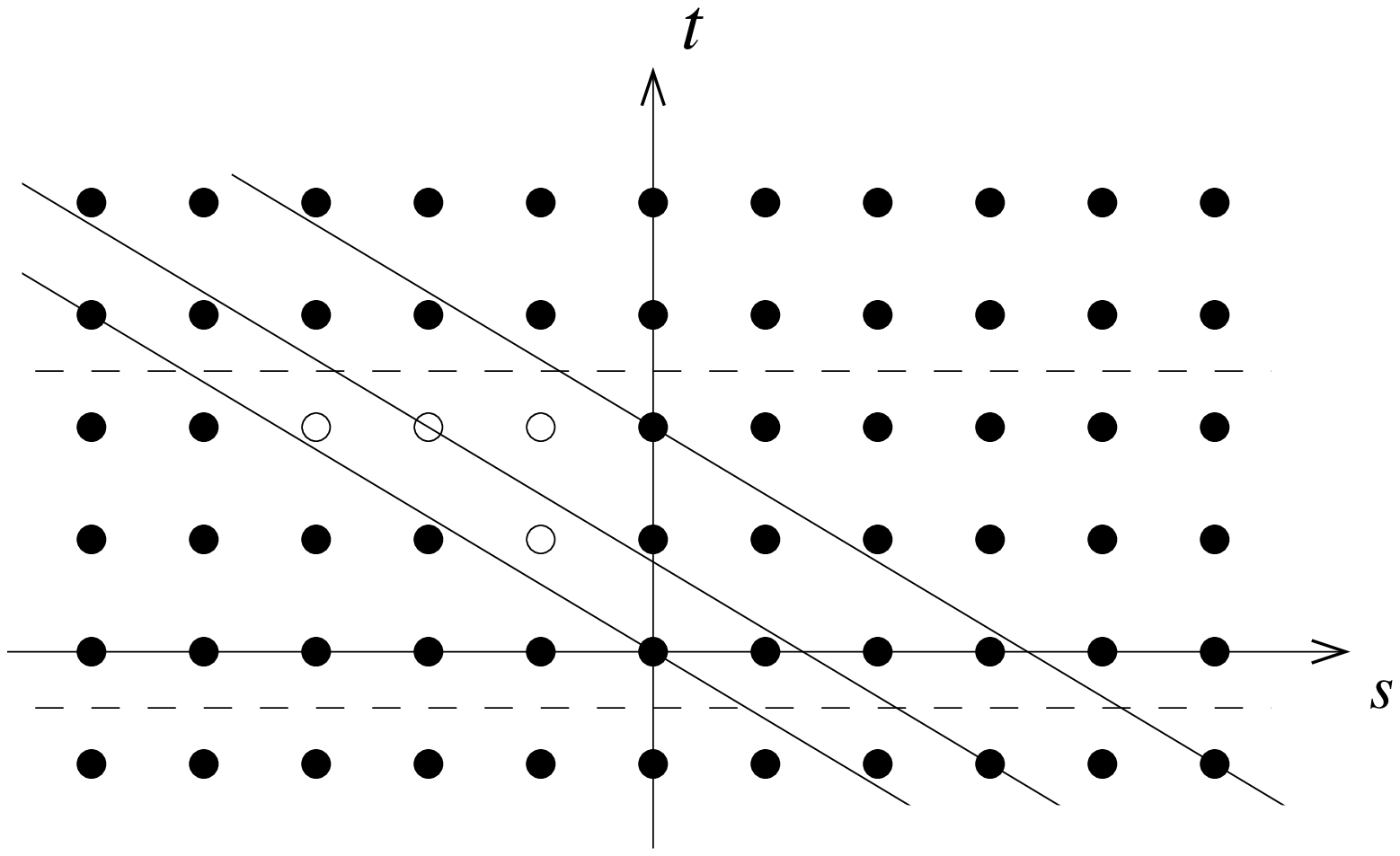}
\\[11pt]
\parbox{0.95\linewidth}{
{\small Figure \ref{frob}: A graphical representation of the
nonrepresentable integers for $\p=3$, $\q=5$, $\NN_{\p
\q}=\{1,2,4,7\}$. The elements of $\NN_{\p \q}$ correspond to the four
unshaded points. Three of the lines $(s,t)=(s_0+\q k,t_0-\p k)$ have
also been shown; each such line contains exactly one point in the
region $0\le t<\p$ between the two dotted horizontal lines. }}
\end{array}
\]
\refstepcounter{figure}
\label{frob}

Substituting back using  (\ref{tildel}) and (\ref{cdelta}), the
allowed values for the conformal weights $\Delta$ precisely
reproduce the Kac table for the minimal model $\CM_{\p\,\q}$:
\eq
\Delta= \Delta_{s,t} = \frac{(\p s -\q t)^2 - (\p-\q)^2}{4 \p \q}~~,~~ 1 \le
t <  \p,~ 1 \le s <\q t/\p~.
\en
It is striking that the full Kac table should emerge from such a
simple consideration of the monodromy properties of the transformed
differential equation (\ref{finale}). Finally, notice that
everything is symmetric in $\p$ and $\q$ so the same result can be
obtained by starting from $M+1=\p/\q$ instead.

Another way to characterise $\NN_{\p \q}$ is through the generating
function
\eq
\pp(z)=\frac{(1-z^\p)(1-z^\q)-(1-z)(1-z^{\p\q})}{(1-z)(1-z^\p)(1-z^\q)}~.
\en
It is straightforward to show that such a rational function is actually
a polynomial, because all the zeros of the denominator are cancelled by
zeros of the numerator; we would like to show that
\eq
\pp(z)=\sum_{2\tilde l+1\in\NN_{\p\,\q}}z^{2\tilde l+1}~.
\label{genera}
\en
To see this, first note that Taylor expanding  $1/(1-z^\p)(1-z^\q)$
yields
\eq
\frac{1}{(1-z^\p)(1-z^\q)}=\sum_{n\in\Rth_{\p\,\q}}c_n\,z^n ~, ~(c_n>0)~.
\en
In order to get rid of the unknown $c_n$'s,  we combine
the trivial identity
\eq
\frac1{1-z^\p}=\sum_{n=0}^\infty z^{n\p\q}\sum_{r=0}^{\q-1}z^{r\p}
\en
with the similar one for $1/(1-z^\q)$ to write
\eq
\frac{1}{(1-z^\p)(1-z^\q)}=\sum_{n=0}^\infty (n+1)\,z^{n\p\q}
\sum_{r=0}^{\q-1} \sum_{s=0}^{\p-1}  z^{r\p+s\q}~.
\en
As a consequence the generating function of representable integers is
\eq
\frac{1-z^{\p\q}}{(1-z^\p)(1-z^\q)}=\sum_{n=0}^\infty z^{n\p\q}
\sum_{r=0}^{\q-1} \sum_{s=0}^{\p-1}  z^{r\p+s\q}=
 \sum_{n\in\Rth_{\p\,\q}}z^n  ~.
\en
Therefore the difference
\eq
\pp(z)\equiv\frac1{1-z}-\frac{1-z^{\p\q}}{(1-z^\p)(1-z^\q)}
\en
is the sought after formula (\ref{genera}).

Before concluding this section we would like to mention that there
is another ODE that can be associated with the same series of minimal
models. This is the so-called $A^{(2)}_2$ description, related to
$\phi_{12}$, $\phi_{21}$ and $\phi_{15}$
perturbations~\cite{Dorey:1999pv}. After a simple change of
variable, the relevant third-order $\phi_{12}$-related ODE can be
cast into the form
\eq
 \left[
\left(\frac{d}{dz} -\frac{\tilde g}{z} \right) \left(\frac{d}{dz}
\right) \left(\frac{d}{dz }
+\frac{ \tilde g}{z} \right)+  ( z^{2\q-3} -
\tilde E z^{\p-3})
\right]y(z,E, \tilde g) =0~.
\label{phi12}
\en
Swapping $\p$ and $\q$
gives the $\phi_{21}$-related ODE, while
replacing
$\p$ with $2\p$ and $\q$ with $\q/2$
yields the $\phi_{15}$ equation.
In all cases, the indicial equation is
\eq
(\lambda + \tilde g) (\lambda -1)(\lambda -(\tilde g+2))=0~.
\en
The zero relative monodromy condition among the three  solutions to
(\ref{phi12}) requires  $(\tilde g+1)$  to be a (positive) integer,
and to avoid  logarithmic  terms we should also  simultaneously impose the
following two conditions
\eq
(\tilde g+1) \notin \{ 2\q t + \p s \}~~,~~2(\tilde g+1) \notin
\{2\q t+ \p s
\}
\label{newcond}
\en
with $s,t=0,1,2,\dots~.$ For $\p$ odd, it is easy to check that
the equations (\ref{newcond}) lead to the  same set of integers as the
$su(2)$-related case discussed above, while only a subset is
recovered for $\p$ even.
For the $\phi_{21}$ related case,
the opposite situation occurs: for $\p$ even the full Kac
table is recovered, while
for $\p$ odd only a subset is found.

\section{Further generalisations and conclusions}

There are many possible
  generalisations of the  above
results. The existence of a simpler version, equation
(\ref{finale}), of the basic
ODE for minimal CFTs is not restricted to  $c<1$ Virasoro models,
but generalises to the higher $su(2)$ coset CFTs
of~\cite{Lukyanov:2006gv} and to the ABCD-related theories
of~\cite{Dorey:2006an}.  The pseudo-differential
equations  listed in \S 3 of~\cite{Dorey:2006an} include
 the minimal models
\eq
\frac{\hat{\mathfrak{g}}_{L} \times
  \hat{\mathfrak{g}}_K}{\hat{\mathfrak{g}}_{L+K}}~,~~~~\mathfrak{g}
=A_n,B_n,C_n,D_n
\label{cosetkl}
\en
at fractional level $L=K\p/(\q-\p)-h^\vee$ with $\q-\p=K u$, and
$u=1,2,\dots$ (using the notation of Appendix 18.B of
\cite{DiFrancesco:1997nk}).
We checked that after simple changes of variable, these equations
reduce to
equations similar in form  to the originals, but for a change in
the `potential', as follows:
\eq
P_K(x)=(x^{h^\vee(\q-\p)/\p K} -E)^K  \longrightarrow  W_{(K,L)}(z)=
z^{\p-h^\vee}
  (z^{(\q-\p)/K}-
\tilde E)^K~.
\en
It is striking that when both  $L$  and $K$ are integer the CFT is
unitary and $W_{(K,L)}(z)$ simplifies further   to
\eq
W_{(K,L)}(z)= z^L (z- \tilde E)^K~.
\label{simpler}
\en
Equation (\ref{simpler}) motivates some simple comments and
speculations. First we observe that  the $K
\leftrightarrow L$ invariance of
(\ref{cosetkl}) manifests itself in (\ref{simpler}) as a  shift
in $z$. As a consequence of this symmetry,  lateral
quantisation problems for the ground-state ODEs and
the associated Stokes multipliers are, up to $\tilde E
\rightarrow -\tilde E$, invariant under the exchange of $L$ and $K$.
However, this symmetry is explicitly broken in equations with  extra
Fuchsian singularities as in (\ref{finale}). A possible remedy is to
treat the points $z=0$ and $z^{(\q-\p)/K}=
\tilde{E}$ more  democratically. For instance, in unitary models  the  symmetry is
globally restored  after the addition of a second singularity  at
$z=\tilde
 E$, ensuring that the set of ODEs for the primary fields in a given
 CFT  is mapped into itself by the
transformation. We suspect that a similar  modification may also
resolve the problem of the missing states  in the $A^{(2)}_2$
example discussed at the end of \S2.

A further
possibility is suggested by equation (\ref{simpler}).
Consider the following multi-parameter generalisation of
(\ref{simpler}):
\eq
W(z,{\bf e})= z \prod_{i=1}^{K+L-1}( z -e_i)
\label{newpot1}
\en
where the  constants $e_i$ ($i=1,2\dots, K+L-1$) are free
parameters.
To keep the discussion brief, we shall restrict attention to
the ground-state equation
for $L+K=3$ and $\mathfrak{g}=su(2)$. Then if $e_1=0$ and $e_2=\tilde E$  the
corresponding ODE  is related to the  tricritical Ising model
$\CM_{4\,5}$, while for $(e_1,e_2)=(\pm
\sqrt{\tilde E} ,\mp \sqrt{\tilde{E}})$ the equation  corresponds
to $\CM_{3 \,5}$. Therefore this simple 2-parameter model
interpolates smoothly between equations associated with
$\CM_{4\,5}$ and $\CM_{3 \,5}$.
This phenomenon has a counterpart in
the homogeneous sine-Gordon model corresponding
to integrable perturbations of the  $\hat{su}(3)_2/U(1)^2$ coset model.
The thermodynamic Bethe ansatz (TBA) equations for this model
\cite{CastroAlvaredo:1999em,Dorey:2004qc}
have two
independent scale parameters $\mu_1$ and $\mu_2$. If one of these
parameters is set to zero, the TBA equations reduce to those
for $\CM_{4\,5}+\phi_{13}$~\cite{Zamolodchikov:1991vh},
while for $\mu_1=\mu_2$ the TBA equations map into a pair of
identical equations for
$\CM_{3\,5}+\phi_{13}$~\cite{Ravanini:1992fi}. This,  and other simple
considerations, suggest that ODEs with multi-parameter potentials of
the form (\ref{newpot1}) may  have an interesting interpretation in
terms of conformal field theory. Much more work will be needed in order
to give this observation a more solid grounding,
but we feel that it will be an interesting direction for
future exploration.
\medskip

\noindent{\bf Acknowledgements --}
We are very grateful to  Junji Suzuki and Edward Frenkel for useful
conversations. We would also like to take this opportunity to
acknowledge the great debt over many years that our whole subject owes
to Aliocha Zamolodchikov, of whose untimely passing we heard
while working on this paper.
PED and TCD thank Torino University and the INFN for
hospitality at the beginning of this project, and PED thanks the Isaac
Newton Institute for hopsitality at the end.
 This project was also partially supported by  grants from the
 Nuffield Foundation, grant number NAL/32601, and from the
 Leverhulme Trust.
%
%
%

%
\end{document}